\documentclass[prl,superscriptaddress,floatfix,showpacs,aps,reprint,letterpaper,nobalancelastpage]{revtex4-1}
\usepackage{graphicx} 
\usepackage{color} 
\usepackage{transparent}
\usepackage{amsmath}
\usepackage{hyperref} 
\usepackage{amssymb}
\usepackage{amsbsy}
\usepackage{url}

\usepackage{multirow}

\newcommand{\ket}[1]{\left\lvert #1 \right\rangle}

\begin{document}
\title{Demonstration of weight-four parity measurements in the surface code architecture}

\author{Maika Takita}
\email{mtakita@us.ibm.com}
\affiliation{IBM T.J. Watson Research Center, Yorktown Heights, NY 10598, USA}
\author{A. D. C\'orcoles}
\affiliation{IBM T.J. Watson Research Center, Yorktown Heights, NY 10598, USA}
\author{Easwar Magesan}
\affiliation{IBM T.J. Watson Research Center, Yorktown Heights, NY 10598, USA}
\author{Baleegh Abdo}
\affiliation{IBM T.J. Watson Research Center, Yorktown Heights, NY 10598, USA}
\author{Markus Brink}
\affiliation{IBM T.J. Watson Research Center, Yorktown Heights, NY 10598, USA}
\author{Andrew Cross}
\affiliation{IBM T.J. Watson Research Center, Yorktown Heights, NY 10598, USA}
\author{Jerry M. Chow}
\affiliation{IBM T.J. Watson Research Center, Yorktown Heights, NY 10598, USA}
\author{Jay M. Gambetta}
\affiliation{IBM T.J. Watson Research Center, Yorktown Heights, NY 10598, USA}

\begin{abstract}
We present parity measurements on a five-qubit lattice with connectivity amenable to the surface code quantum error correction architecture. Using all-microwave controls of superconducting qubits coupled via resonators, we encode the parities of four data qubit states in either the $X$- or the $Z$-basis. Given the connectivity of the lattice, we perform full characterization of the static $Z$-interactions within the set of five qubits, as well as dynamical $Z$-interactions brought along by single- and two-qubit microwave drives. The parity measurements are significantly improved by modifying the microwave two-qubit gates to dynamically remove non-ideal $Z$ errors.
\end{abstract}

\maketitle

The fragile nature of quantum information means that the success of large-scale quantum computing hinges upon the successful implementation of quantum error correction (QEC) on physical qubit systems. Typically QEC protocols function through encoding of physical qubit information onto larger subspaces, which are subsequently protected against particular quantum errors~\cite{Shor1995,kitaev_fault-tolerant_2003}. Amongst the many proposed QEC codes, the topological surface code~\cite{Bravyi:1998, Raussendorf:2007} has gathered a large amount of interest by experimental implementations~\cite{Chow:2014,Kelly:2015} due to its use of short-range nearest-neighbor interactions between physical qubits and its relatively high error thresholds.

Building up a physical quantum network with the complete functionality of the surface code brings along a number of experimental challenges, some of which have yet to be explored. However, in the particular case of superconducting qubits, recent advances in coherence times~\cite{Paik2011, Rigetti:2012,chang_improved_2013} and in the understanding of environmental constraints~\cite{Barends:2011,Corcoles:2011} have triggered important experimental demonstrations on increasingly larger systems, including correction of bit-flip errors on linear qubit arrays~\cite{Kelly:2015,Riste:2015}, the detection of arbitrary quantum errors~\cite{Corcoles:2015}, and state preservation via encoding in cavity coherent states~\cite{Ofek:2016}. With gate fidelities continuing to improve~\cite{Barends:2014,Sheldon:2016}, it becomes critical to demonstrate the ability to perform these operations in systems with the degree of connectivity required by the surface code. Furthermore, exploring higher-order errors in such non-trivially arranged networks of qubits are necessary for outlining the proper route towards larger numbers of interconnected qubits for QEC.

In this Letter we demonstrate a plaquette of the surface code QEC protocol with an interconnected network of five superconducting transmon qubits. This network consists of four data qubits each explicitly coupled to a single syndrome qubit, through which single-shot high-fidelity readout is used to measure weight-four checks of both the bit-flip and phase-flip data qubit parity. The geometrical arrangement of the network permits a systematic calibration of crosstalk noise within the plaquette, and we specifically look for errors in non-participating, or ``spectator" qubits, during two-qubit gates. To make the plaquette operation free of these crosstalk errors, we develop and benchmark a new decoupling sequence for our two-qubit gates which involve action on the spectator qubits as well, and show the ability to maintain gate-fidelities close to the limits given by coherence. The resulting probability of obtaining the right parity in the plaquette is subsequently measured to be 0.774 (0.795), with a standard deviation of 0.013, for a weight-four $ZZZZ$-($XXXX$-) parity measurement. 

\begin{figure}[tp]
	\includegraphics[width=0.47\textwidth]{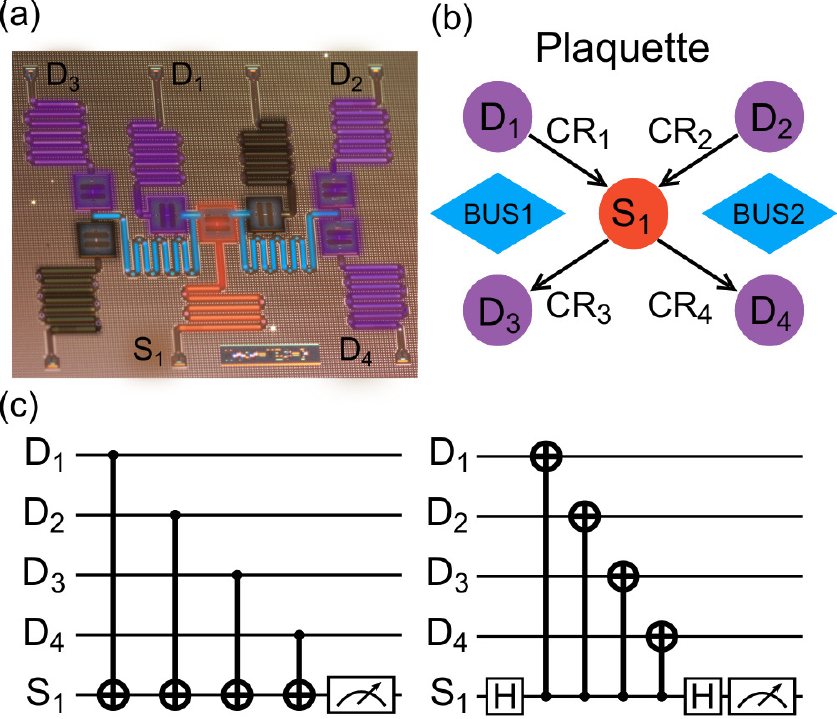}
	\caption{(color online) (a) False-colored picture of a 7-qubit lattice. The five specific qubits used for the plaquette experiment are highlighted and labeled as data qubits ($\text{D}_i$, $i \in [1,4]$) and syndrome qubit $S_1$. (b) Cartoon representation of a plaquette. The arrows represent the cross-resonance two-qubit gate directions between the syndrome qubit and the four data qubits, with the convention of pointing from control to target qubit. The two quantum bus resonators, bus 1 and bus 2, are measured to be $\omega_{B1}/2\pi = 6.562$ GHz and $\omega_{B2}/2\pi = 6.810$ GHz respectively. (c) The $ZZZZ$ and $XXXX$-parity measurement quantum circuits. \label{fig:1}}
\end{figure}

The five-qubit plaquette (for schematic see Fig.~\ref{fig:1}(b)) is contained within a larger lattice of seven fixed-frequency superconducting transmon qubits shown in Figure \ref{fig:1}(a). In this particular arrangement, two separate microwave cavity buses independently couple to four qubits. The central qubit, which serves as the syndrome qubit of the plaquette, is coupled to both buses. Of the three remaining qubits coupled to each bus, any two of the qubits can be used as data qubits in the central plaquette. The fabrication process of the device is described in previous work~\cite{Chow:2014,Corcoles:2015} and the design of the transmon is specifically chosen to minimize the participation ratios of specific lossy interfaces~\cite{Wang:2009,Wenner:2011,Wang:2015a,Dial:2015}. Qubit transition frequencies ($\omega_{01}/2\pi$) of the four data qubits, D$_1$-D$_4$, are $\{4.79098, 4.80196, 4.89785, 4.94908\}$ GHz, and the syndrome qubit, S$_1$, is 4.65808 GHz. The measured parameters of the transmon qubits and associated individual readout resonators are given in the Supplemental Material~\cite{SupplementalMaterial}. Any qubit state can be individually measured via the independent readout resonators with signals which are then amplified via Josephson Parametric Converters (JPCs)~\cite{Bergeal:2010,Abdo:2011aa}.

Given the connectivity of the five-qubit plaquette, and the particular arrangement of fixed frequencies, it is crucial to verify the operability of single-qubit operations both independently and at the same time. Using Clifford randomized benchmarking (RB)~\cite{Magesan:2011}, we find all single-qubit gate fidelities to be in excess of 0.998. The gates are 50 ns long and the pulse-shapes and calibration sequences are described in previous work~\cite{Corcoles:2015}. With an arrangement of interconnected fixed-frequency qubits, it is critical to characterize addressability errors for the single-qubit gates. Employing the technique of  simultaneous RB~\cite{Gambetta:2012}, we find no significant reduction in any of the single-qubit gate fidelities, when all five qubits are benchmarked at the same time, thus bounding addressability errors to $<$0.00083. A complete summary of these single-qubit benchmarking results is given in the Supplemental Material~\cite{SupplementalMaterial}.

\begin{figure}[htp]
	\includegraphics[width=0.47\textwidth]{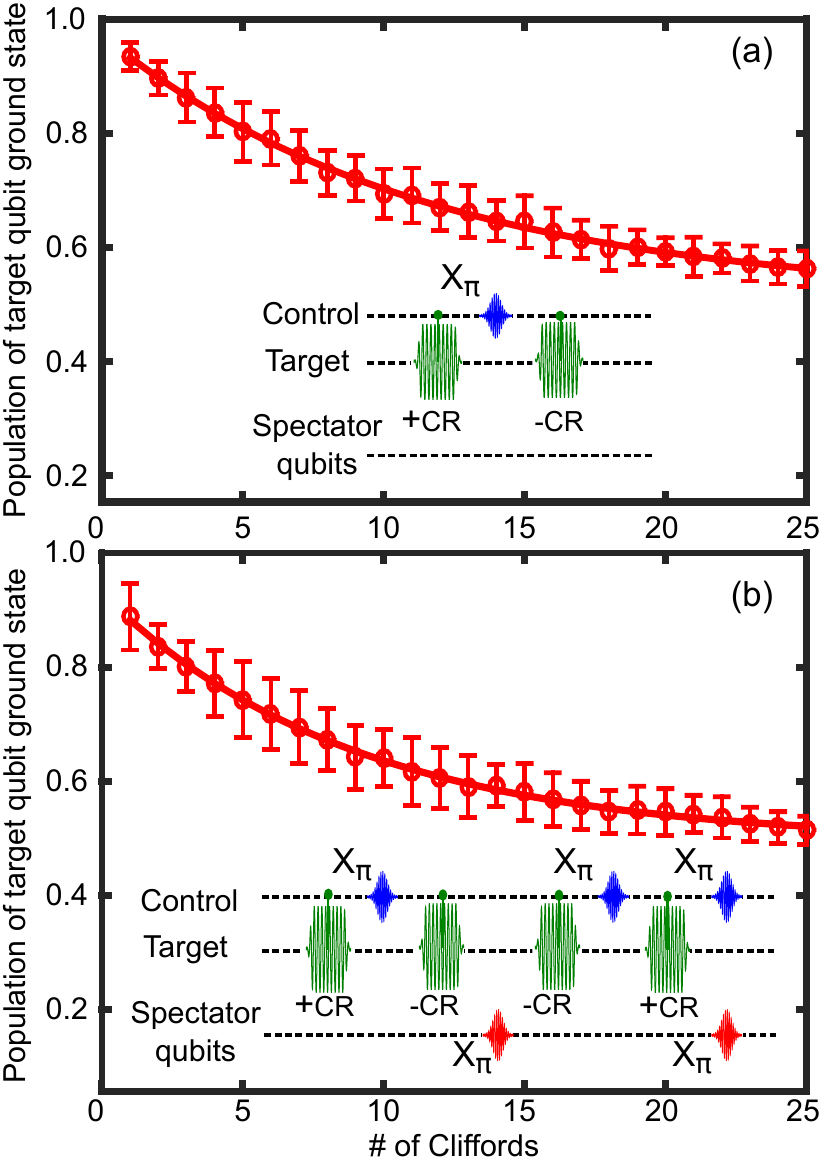}
	\caption{(color online) Two-qubit gate ($CR_2$) randomized benchmarking results with (a) $\text{ECR}_{\text{2-pulse}}$ Clifford generator and (b) a $\text{ECR}_{\text{4-pulse}}$ Clifford generator using the cross-resonance interaction with $D_2$ as the control and $S_1$ as the target qubit. Inset of the figures show the different ECR gate pulse sequence decompositions. We show the decay of the ground state population of the target qubit. Data for the control qubit gives the same RB figure within our confidence interval. \label{fig:2}}
\end{figure}

At the core of the plaquette operation are two-qubit controlled-NOT (CNOT) gates between each of the four data qubits and the syndrome. Given the resonator bus configuration of our device and the fixed-frequencies of the qubits, a microwave-based cross-resonance interaction~\cite{Paraoanu2006,Rigetti2010,Chow:2011} is supported between each of the two data qubits in either bus to the central syndrome qubit. A two-qubit echo cross-resonance (ECR) gate, $\text{ECR}_{\text{2-pulse}} = ZX_{90} - XI$, (which is a CNOT up to single-qubit rotations) can be constructed from the base cross-resonance interaction through a simple control-qubit echo decoupling~\cite{Corcoles:2013} sequence [see inset of Fig.~\ref{fig:2}(a)], involving two opposite amplitude cross-resonance (CR) drives. As $\text{ECR}_{\text{2-pulse}} $ is also a generator for two-qubit Clifford gates, we can characterize each of the four cross-resonance gates through Clifford RB, finding all two-qubit ECR fidelities to be in excess of 0.947 with total gate lengths between 340~ns to 1010~ns long. Two-qubit benchmarking results are summarized in Table~\ref{table:2}. Each RB experiment reported in this work, single- and two-qubit, was obtained over 30 different random sequences.

\begin{table}[ht]
	\begin{ruledtabular}
		\begin{tabular}{|c|c|c|c|c|}
			\hline
			&$CR_1$ 	   &$CR_2$ 		&$CR_3$ 		&$CR_4$ 
			\\ \hline
			$\text{ECR}_{\text{2-pulse}}$ 		&$0.9637$  	&0.9572 	&0.9523 	&0.9469 \\
			fidelity 		&$\pm0.0014$ 	&$\pm0.0014$ 	&$\pm0.0001$ 	&$\pm0.0007$ \\
			(gate length, ns) 	&(660) 		&(340)	&(720)	&(1010)
			\\ \hline
			$\text{ECR}_{\text{4-pulse}}$  		&0.9405  		&0.9458 	&0.9469 	&0.9384\\
			fidelity	&$\pm0.0011$ 	&$\pm0.0016$ 	&$\pm0.0015$ 	&$\pm0.0013$  \\ 
			(gate length, ns)	&(740) 		&(580)	&(820)	&(940) \\
			\hline \hline			
			
		\end{tabular}
	\end{ruledtabular}
	\caption{\label{table:2} \textbf{Two-qubit gate fidelities.} Randomized benchmarking results and total gate times for ECR gates with 2-pulse cross-resonance ($\text{ECR}_{\text{2-pulse}}$) and 4-pulse decomposition cross-resonance ($\text{ECR}_{\text{4-pulse}}$) gates. Powers used for $\text{ECR}_{\text{4-pulse}}$ are different from the powers used for $\text{ECR}_{\text{2-pulse}}$. See insets of Fig.~\ref{fig:2} for depiction of the two-qubit gate pulse sequences.}
\end{table}

To study the weight-four $ZZZZ$-($XXXX$-) parity measurement, we first prepare each of the four data qubits into either a $\ket{0}$ ($\ket{+}$) or a $\ket{1}$ ($\ket{-}$) state to initialize into $2^4 = 16$ different states. For each one of the sixteen states, we ran four pairs of CNOT gates sequentially to measure the weight four parity. The full measurement circuit is shown in Fig.~\ref{fig:1}(c). Using the readout calibration method described in previous work~\cite{Ryan:2015,Corcoles:2015}, a threshold value is assigned from 20,000 single shot measurements with 2.424$\mu$s integration time for an assignment fidelity of 0.925. The mean of sixteen probabilities for measuring the correct $ZZZZ$-($XXXX$-) parity is 0.701 (0.570) with standard deviation of 0.032 (0.009) over 20,000 repeated measurements. While the result from $ZZZZ$-parity checks agree with the expected fidelity calculated from the gate fidelities obtained from RB and readout assignment fidelities, $XXXX$-parity checks show much lower fidelity than expected which indicates the presence of $Z$-errors in the parity check operations.

In order to gain some insight on the origin of these $Z$-errors we perform a series of Ramsey interferometry experiments on each of the qubits in the presence of cross-resonance driving pulses and for different $Z$-eigenstates of the remaining qubits. These experiments also allow us to differentiate between static $Z$-interactions spanning our qubit system and dynamic $Z$-interactions activated by microwave control tones. 

The evolution of the state of the three spectator qubits and the control qubit during the CR drive can be described by the Hamiltonian $\mathcal{H}=\sum_{ijkl=0,1} \eta_{ijkl}|ijkl\rangle \langle ijkl|$, where we will assign the first index to the control qubit. This Hamiltonian can also be written as $\mathcal{H}=\sum_{ijkl=0,1} \alpha_{ijkl}Z^iZ^jZ^kZ^l/2$, where $Z^0 = I$ and $Z^1 = Z$. Thus, the different $\alpha_{ijkl}$, which we can recover from the $\eta_{ijkl}$ obtained from the interferometry experiments, give the strength of the $Z-$interactions (see Supplemental Material~\cite{SupplementalMaterial} for details).

Our maximum Ramsey delay time was $\sim 33$ $\mu$s, which gives us a frequency resolution on the order of that determined by the coherence of our system. Measuring interactions below $\sim 100$ kHz becomes challenging. We can, however, determine the largest source of $Z-$errors in our operations and devise a mitigating strategy.

\begin{figure}[t]
	\includegraphics[width=0.5\textwidth]{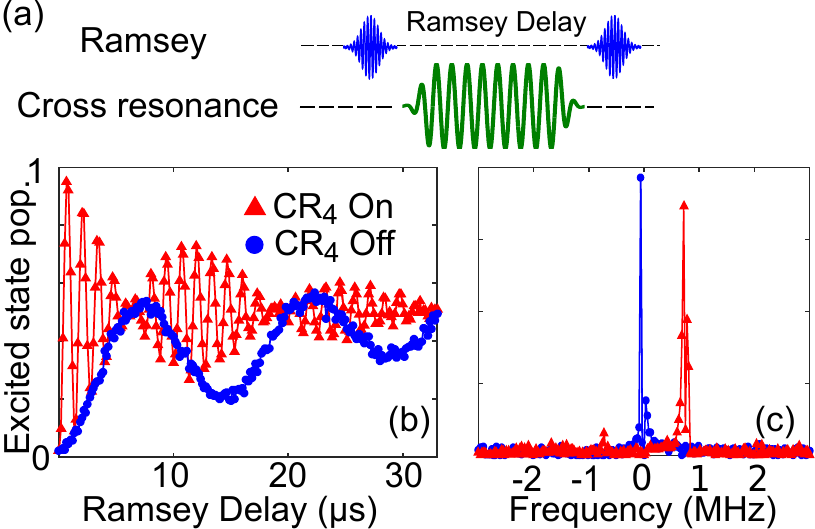}
	\caption{(color online) (a) Pulse sequence for the CR Ramsey experiments. The CR resonance tone spans the entirety of the Ramsey delay time and is applied with the same amplitude used in the parity check operations. Time domain (b) and Fourier Transform (c) of Ramsey interferometry on $D_3$ with $CR_4$ on (triangles) and off (circles) for the case where all qubits other than $D_3$ and the target $S_1$ are kept at the ground state. The CR Ramsey experiment, along with equivalent ones for all eight $D_1$, $D_2$ and $D_4$ states, translate into the $IIIZ$ and $ZIIZ$ strengths shown in Table~\ref{table:CRZZs} for $CR_4$.  We attribute the slower oscillations in the no CR Ramsey experiment, around 30 kHz, to charge noise affecting $D_3$. \label{fig:3}}
\end{figure}

Table \ref{table:CRZZs} shows the parameters $\alpha_{ijkl}$ obtained from our interferometry experiments. The effect of $CR_4$ on the Ramsey data from $D_3$ is shown in Fig. \ref{fig:3}, both in time domain [Fig.~\ref{fig:3}(b)] and Fourier space [Fig.~\ref{fig:3}(c)]. Fig.~\ref{fig:3}(a) shows the pulse sequence used to perform the Ramsey experiments. 

The presence of $Z-$errors in our architecture is a natural, albeit hitherto relatively unexplored, consequence of the CR Hamiltonian. The bus resonators coupling the qubits in our device allows for a significant degree of control over the $Z-$interactions necessary to implement entangling operations within our gate scheme. The presence of spurious $Z-$terms, however, cannot be completely cancelled by microwave design and require of decoupling techniques as described below. 

All four CR gates have a significant effect in the form of $Z-$interactions on one or more of the spectator qubits. These interactions ($IZ$) are also in some cases dependent on the state of the control qubit ($ZZ$). Double $Z-$errors involving two or more spectator qubits appear below our sensitivity limit of $\sim$ 100 kHz. Whereas these errors can be important in experiments with a larger scope, they do not seem to be the main source of infidelity in the presented plaquette experiments. A careful and thorough study of higher order $Z-$interactions in a system with comparable connectivity to the device here presented will be the topic of future work.

Other $Z-$interactions, including static $ZZ$ between the qubits and $Z-$interactions activated by single qubit operations in neighboring qubits, are largely below our detection limit. 
\begin{table}
	\begin{center}
		\begin{tabular}{r|c|c|c|c|}
	\cline{2-5}
	& $ CR_1 $			& $ CR_2 $		& $ CR_3 $		& $ CR_4 $ 	\\						 
	& $ D_1D_2D_3D_4 $	&  $ D_2D_1D_3D_4 $	& $ S_1D_1D_2D_4 $	& $ S_1D_1D_2D_3 $ \\
    & (kHz) & (kHz) & (kHz) & (kHz)				\\ \cline{1-5}
	\multicolumn{1}{|c|}{$IIIZ$} 	&	-298			&	-688			&	-178			&	640			\\ \cline{1-5}
	\multicolumn{1}{|c|}{$IIZI$} 	&	-348			&	$\epsilon$			&	130			&	$\epsilon$			\\  \cline{1-5}
	\multicolumn{1}{|c|}{$IZII$} 	&	-129			&	-140			&	$\epsilon$			&	$\epsilon$			\\  \cline{1-5}
	\multicolumn{1}{|c|}{$ZIIZ$} 	&	$\epsilon$			&	$\epsilon$			&	113			&	105			\\ \cline{1-5}
	\multicolumn{1}{|c|}{$ZIZI$} 	&	$\epsilon$			&	-129			&	$\epsilon$			&	$\epsilon$			\\  \cline{1-5}	
		\end{tabular}
	\end{center}
	\caption{\label{table:CRZZs} \textbf{CR-activated $Z$ interactions} Approximate $Z$-strengths on spectator qubits brought on by CR tones, in kHz. $\epsilon$ indicates a $Z-$strength below our sensitivity limit of $\sim 100$ kHz and only terms with values above this limit from any of the cross-resonance tones are shown.}{\tiny }
\end{table}

Our preliminary implementation of the two-qubit gate, using $\text{ECR}_{\text{2-pulse}}$, initially developed in Ref. \onlinecite{Corcoles:2013}, did make use of echoing sequences to echo away $ZI$-interaction on the control qubit as well as the $IX$-interaction on the target. However, the $IZ-$ and $ZZ-$errors between the control qubit of the CR and the spectator qubits created by CR pulses cannot be echoed away using the two-pulse CR gate. In order to cancel those terms we employ a more complex pulse sequence consisting of two primitive two-pulse $\text{ECR}_{\text{2-pulse}}$ gates with a $\pi$ rotation on each spectator qubit between them [see Fig.~\ref{fig:2}(b) inset]. This new four-pulse two-qubit gate (which we now call $\text{ECR}_{\text{4-pulse}}$ can be similarly characterized using RB. The results are also given in Table~\ref{table:2} below the RB results for $\text{ECR}_{\text{2-pulse}}$ [see Fig.~\ref{fig:2}(b) with a representative RB result shown]. Note that during the two-qubit gates, all three other spectator qubits are being echoed.  

\begin{figure}[htp]
	\includegraphics[width=0.47\textwidth]{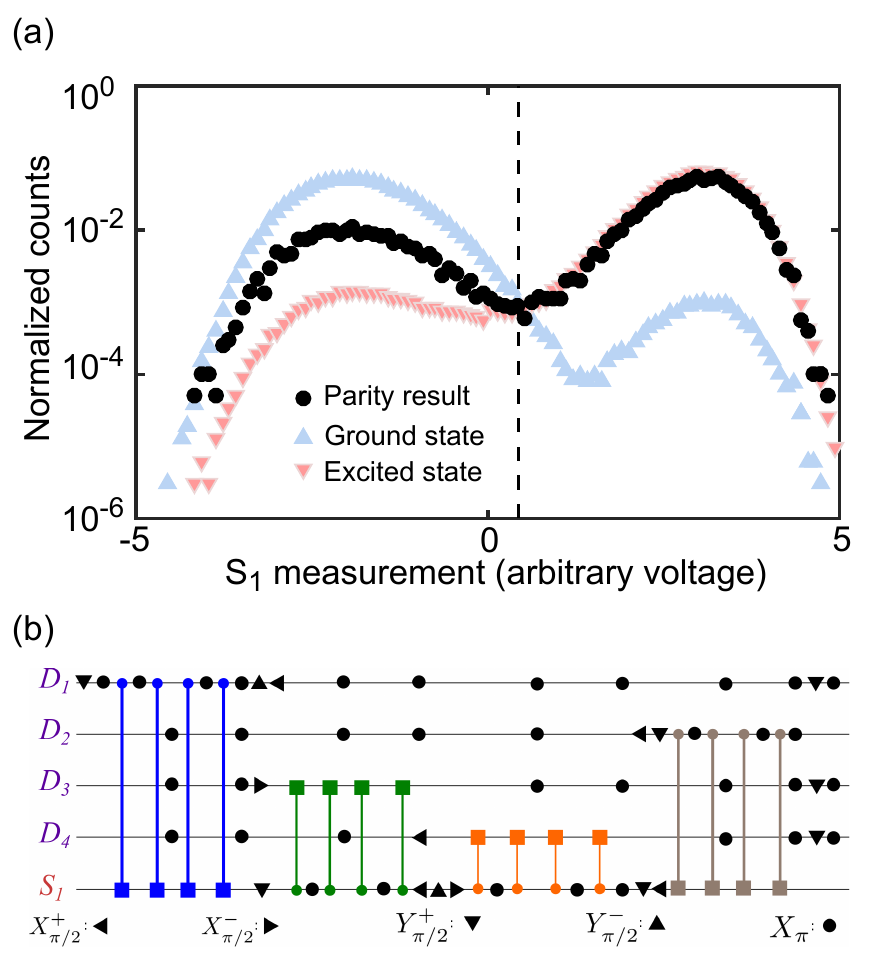}
	\caption{(color online) (a) Histogram of ground (triangles up) and excited (triangles down) state calibrations and $XXXX$-parity check result (circles) when the data qubits ($D_1D_2D_3D_4$) are prepared into the $\ket{--+-}$ state. The dashed line is the threshold used to determine the state of the syndrome qubit, obtained from calibration pulses, where assignment fidelity is calculated to be 97.4\%. In this parity check experiment, 79.4\% of the results lies above the threshold line (P(1)). (b) Complete gate sequence used to run the $XXXX$ parity checks. The pulses that prepare the data qubits in each basis state are not shown. Vertical lines represent each of the cross-resonance pulses from which our four-pulse CNOT primitives are constructed, with squares placed on the target qubit in each case.\label{fig:4}}
\end{figure}

To run the full-plaquette experiment again with this $\text{ECR}_{\text{4-pulse}}$ gate we have optimized the readout chain by adding microwave switches to the single qubit and cross-resonance tones to mitigate mixer leakage. This improved the assignment fidelity to 0.974 with 1.248$\mu$s integration time. With the $\text{ECR}_{\text{4-pulse}}$  gate fidelities and assignment fidelity, we expect both $ZZZZ$ and $XXXX$-parity checks to give the correct parity result with probability 0.76. We find, however, that the correct parity is obtained on average with probability 0.774 and 0.795 -with standard deviation of 0.013- for $ZZZZ$ and $XXXX$, respectively. Full results of the weight-four checks are given in Table~\ref{table:3}. Note that multiple errors during the execution of the plaquette can result in the same parity as an error-free experiment, to give an overall fidelity higher than the combined measured fidelity of the constituting gates. The confidence bounds given in Table~\ref{table:3} for P(0) and P(1) are obtained from the probabilities of measuring the wrong state ($|0\rangle$ when preparing $|1\rangle$ and $|1\rangle$ when preparing $|0\rangle$) in the calibration pulses used to determine the assignment fidelity (triangles in Fig.~\ref{fig:4}). 

On a different five qubit device~\cite{QExpwebpage} we have repeated the plaquette experiments with similar results.

\begin{table}[htp]
	\centering
	
	\begin{minipage}{.24\textwidth}
		\begin{tabular}{|c|c|c|}
			\hline
			$\ket{D_1D_2D_3D_4}$ 	 & \:\:$P_C$\:\:   & 	 \:\:SE\:\: \\ \hline
			$\ket{0\: 0\: 0\: 0}$	& 0.790				& 0.0029		  	\\ \hline
			$\ket{0\: 0\: 0\: 1}$	& 0.782			 	& 0.0029			\\ \hline
			$\ket{0\: 0\: 1\: 0}$	& 0.782			 	& 0.0029			\\ \hline
			$\ket{0\: 0\: 1\: 1}$	& 0.758				& 0.0030		  	\\ \hline
			$\ket{0\: 1\: 0\: 0}$	& 0.775			 	& 0.0030	\\ \hline
			$\ket{0\: 1\: 0\: 1}$	& 0.764				& 0.0030			  	\\ \hline
			$\ket{0\: 1\: 1\: 0}$	& 0.762				& 0.0030			  	\\ \hline
			$\ket{0\: 1\: 1\: 1}$	& 0.746			 	& 0.0032	\\ \hline
			$\ket{1\: 0\: 0\: 0}$	& 0.771			 	& 0.0030	\\ \hline
			$\ket{1\: 0\: 0\: 1}$	& 0.789				& 0.0029			  	\\ \hline
			$\ket{1\: 0\: 1\: 0}$	& 0.768				& 0.0030			  	\\ \hline
			$\ket{1\: 0\: 1\: 1}$	& 0.765			 	& 0.0030	\\ \hline
			$\ket{1\: 1\: 0\: 0}$	& 0.782 			& 0.0029			  	\\ \hline
			$\ket{1\: 1\: 0\: 1}$	& 0.786			 	& 0.0029	\\ \hline
			$\ket{1\: 1\: 1\: 0}$	& 0.788			 	& 0.0029	\\ \hline
			$\ket{1\: 1\: 1\: 1}$	& 0.778				& 0.0029			  	\\ \hline			
		\end{tabular}
	\end{minipage}\hfill
	\begin{minipage}{.24\textwidth}
		\begin{tabular}{|c|c|c|}
			\hline
			$\ket{D_1D_2D_3D_4}$ 	 & \:\:$P_C$\:\:   & 	 \:\:SE\:\:\\ \hline
			$\ket{++++}$	& 0.790				& 0.0029			  	\\ \hline
			$\ket{+++-}$	& 0.805			 	& 0.0028					\\ \hline
			$\ket{++-+}$	& 0.797			 	& 0.0028	\\ \hline
			$\ket{++--}$	& 0.773				& 0.0030			  	\\ \hline
			$\ket{+-++}$	& 0.784			 	& 0.0029	\\ \hline
			$\ket{+-+-}$	& 0.806				& 0.0028			  	\\ \hline
			$\ket{+--+}$	& 0.808				& 0.0028			  	\\ \hline
			$\ket{+---}$	& 0.784			 	& 0.0029	\\ \hline
			$\ket{-+++}$	& 0.796			 	& 0.0028	\\ \hline
			$\ket{-++-}$	& 0.772				& 0.0030			  	\\ \hline
			$\ket{-+-+}$	& 0.795				& 0.0029			  	\\ \hline
			$\ket{-+--}$	& 0.797			 	& 0.0028	\\ \hline
			$\ket{--++}$	& 0.822				& 0.0027			  	\\ \hline
			$\ket{--+-}$	& 0.794			 	& 0.0029	\\ \hline
			$\ket{---+}$	& 0.792			 	& 0.0029	\\ \hline
			$\ket{----}$	& 0.806				& 0.0028			  	\\ \hline		
		\end{tabular}
	\end{minipage}\hfill
	\caption{\label{table:3} \textbf{Parity-Check Results.} Probability $P_C$ of measuring the correct parity of the data qubit input states after running a $ZZZZ$ or a $XXXX$-parity check (see Fig.~\ref{fig:1}(c)). The statistic error (SE) in each case is estimated from the variance ($\sqrt{P_C(1-P_C)}$) and the number of shots (20,000 in each case). We can associate a systematic error to our results from the probability of obtaining $\ket{0}$ when preparing $\ket{1}$ (0.025) and the probability of obtaining $\ket{1}$ when preparing $\ket{0}$ (0.016).}	
\end{table}

The demonstrated plaquette experiments constitute a key stepping stone in our advances towards fault-tolerant implementations of error correcting schemes. We have shown that within a connectivity at the level required to run the surface code, high order $Z$-interactions (involving qubits not intervening in the cross-resonance gate) are not a serious source of error at the fidelity attained by our two-qubit unitaries. As our understanding of the cross-resonance Hamiltonian improves for highly connected systems, we will be able to devise new methods to characterize and compensate for –or exploit- its higher-order interactions in order to bring the fidelity of our parity checks up beyond fault-tolerance thresholds.

\begin{acknowledgments}
We acknowledge experimental contributions from Jim Rozen and Matthias Steffen. We acknowledge support from IARPA under contract W911NF-10-1-0324. All statements of fact, opinion or conclusions contained herein are those of the authors and should not be construed as representing the official views or policies of the U.S. Government.
\end{acknowledgments}
\bibstyle{misc}
\bibliography{qc_supcon}

\clearpage

\onecolumngrid
\vspace{\columnsep}

	\section{Supplementary material for `Demonstration of weight-four parity measurements in the surface code architecture'}

\subsection{System characterization}

\begin{table}[h]
	\begin{tabular}{|c|c|c|c|c|c|}
		\hline
		Qubit 		           & $\text{D}_1$ 	    & $\text{D}_2$ 	    & $\text{D}_3$ 	    & $\text{D}_4$ 	    & $\text{S}_1$	\\ \hline
		$\omega_{01}/2\pi$ (GHz) &	4.79098		&	4.80196		&	4.89785		&	4.94908		& 4.65808
		\\ \hline
		$T_1 \pm std$ ($\mu$s)   & $35.1 \pm1.3$  & $46.9\pm3.6$  & $56.1\pm2.7$  & $48.5\pm2.5$   & $54.3\pm2.8$
		\\ \hline
		$T_2 \pm std$ ($\mu$s)  & $40.8 \pm2.1$  & $31.4\pm4.1$  & $44.7\pm2.7$  & $35.5\pm1.7$  &   $44.0\pm2.4$
		\\ \hline
		$\omega_{R}/2\pi$ (GHz)   &	6.5892		& 6.6957		&	6.5268	& 6.5613		& 6.7164 
		\\ \hline
		$2\chi/2\pi$ (kHz) & 460  &  470 & 630  & 630  & 450
		\\ \hline
		$\kappa/2\pi$ (kHz) & 560  & 450  & 630  & 500  &  560 
		
		\\ \hline \hline 			
		Single Qubit & 0.99928  	& 0.99923 	& 0.99925 	& 0.99909 	& 0.99841	\\
		RB			& $\pm0.00002$ & $\pm0.00004$ & $\pm0.00002$ & $\pm0.00002$ & $\pm0.00004$ \\ \hline
		Simultaneous & 0.99845  	& 0.99919 	& 0.99917 	& 0.99904 	& 0.99807	\\
		RB			& $\pm0.00004$ & $\pm0.00007$ & $\pm0.00003$ & $\pm0.00002$ & $\pm0.00004$ \\ \hline
		
	\end{tabular}
	\caption{\label{table:1} \textbf{Qubit and readout characterization.} Qubit transitions ($\omega_{01}/2\pi$), relaxation times ($T_1$), coherence times ($T_2^{echo}$), readout resonator frequencies ($\omega_{R}/2\pi$), dispersive shifts (2$\chi$), line widths ($\kappa$), and individual and simultaneous single-qubit randomized benchmarking results. Single qubit gates are all 50 ns long. Anharmonicity of the qubits are all around 340 MHz. }	
\end{table}

\subsection{Extracting $Z$-interaction strengths from Ramsey measurements}

In order to get a measure of the strength of the $Z$-interactions brought by the cross-resonance (CR) drive, we perform a series of Ramsey interferometry measurements on each spectator qubit, with the rest of the spectators and control qubit in all possible $Z-$basis states.

If we write the four-qubit Hamiltonian (control qubit plus three spectators) brought up by the application of the CR drive as $\mathcal{H} = \sum_{ijkl=0,1} \eta_{ijkl} |ijkl\rangle \langle ijkl|$, we can see that
\begin{eqnarray*}
	\mathcal{H} &=& \sum_{ijk=0,1} \eta_{ijk0} |ijk0\rangle \langle ijk0|+\sum_{ijk=0,1} \eta_{ijk1} |ijk1\rangle \langle ijk1|=\sum_{ijk=0,1} \eta_{ijk} |ijk\rangle \langle ijk|\otimes \frac{I+Z}{2} +\sum_{ijk=0,1} \eta_{ijk} |ijk\rangle \langle ijk|  \otimes \frac{I-Z}{2} \\
	&=& \frac{\eta_{ijk0}+\eta_{ijk1}}{2} |ijk\rangle \langle ijk| \otimes I + \frac{\eta_{ijk0}-\eta_{ijk1}}{2} |ijk\rangle \langle ijk| \otimes Z
\end{eqnarray*}

Here we associate the first index always to the CR control qubit. The quantity  we extract from our Ramsey experiments for each vaule of $|ijk\rangle$ is $(\eta_{ijk0}-\eta_{ijk1})$ (note that we can not perform Ramsey interferometry on the qubits partaking the CR gate in each case). 

Writing the above four-qubit Hamiltonian in terms of $Z$ operators, we have

\begin{equation*}
\mathcal{H} = \sum_{ijkl=0,1} \alpha_{ijkl}Z^iZ^jZ^kZ^l/2
\end{equation*}
where $Z^0 = I$ and $Z^1=Z$. Our goal is to experimentally determine the numbers $\alpha_{ijkl}$. The factors $\eta_{ijkl}$ and $\alpha_{ijkl}$ are related by the expression

\begin{equation}\label{equation1}
\alpha_{ijkl} = \frac{H^{16\times 16}}{8} \eta_{ijkl}
\end{equation}
where $H^{16 \times 16}$ is the Hadamard matrix of order 16. Now, if we define $\zeta_{ijk\blacktriangle} = \eta_{ijk0}-\eta_{ijk1}$, there is a simple $24\times 16$ matrix $\mathcal{B}$ such that $\bar{\zeta} = \mathcal{B} \bar{\eta}$, where $\bar{\zeta}$ is a $24 \times 1$ column matrix (note that $\zeta_{\blacktriangle ijk}$ is not part of $\bar{\zeta}$ as we do not perform Ramsey experiments on the CR control qubit) and $\bar{\eta}$ is a $16 \times 1$ column matrix of all the $\eta_{ijkl}$.

We can obtain the pseudo-inverse of the matrix $\mathcal{B}$ to relate $\bar{\eta}$ to $\bar{\zeta}$ and, substituting on Eq.\ref{equation1} we obtain

\begin{equation}
\bar{\alpha} = \frac{H^{16\times 16}}{8} \mathcal{B}^{-1} \bar{\zeta}
\end{equation}

We show the values obtained for all $\bar{\zeta}$ in Table \ref{table:STable1}.

\begin{table}[h]
	\begin{center}
		\begin{tabular}{r|c|c|c|c|}
			\cline{2-5}
			& $ CR_1 $			& $ CR_2 $		& $ CR_3 $		& $ CR_4 $ 	\\						 
			& $ D_1D_2D_3D_4 $	&  $ D_2D_1D_3D_4 $	& $ S_1D_1D_2D_4 $	& $ S_1D_1D_2D_3 $ \\
			& (kHz) & (kHz) & (kHz) & (kHz)				\\ \cline{1-5}
			\multicolumn{1}{|c|}{$\zeta_{000\blacktriangle}$} 	&	-320			&	-710			&	-50			&	790			\\ \cline{1-5}
			\multicolumn{1}{|c|}{$\zeta_{001\blacktriangle}$} 	&	-290			&	-680			&	-80			&	730			\\  \cline{1-5}
			\multicolumn{1}{|c|}{$\zeta_{010\blacktriangle}$} 	&	-320			&	-680			&	-50			&	730			\\  \cline{1-5}
			\multicolumn{1}{|c|}{$\zeta_{011\blacktriangle}$} 	&	-290			&	-680			&	-80			&	730			\\ \cline{1-5}
			\multicolumn{1}{|c|}{$\zeta_{100\blacktriangle}$} 	&	-290			&	-710			&	-290			&	520			\\  \cline{1-5}	
			\multicolumn{1}{|c|}{$\zeta_{101\blacktriangle}$} 	&	-290			&	-680			&	-290			&	580			\\ \cline{1-5}
			\multicolumn{1}{|c|}{$\zeta_{110\blacktriangle}$} 	&	-290			&	-680			&	-290			&	520			\\  \cline{1-5}
			\multicolumn{1}{|c|}{$\zeta_{111\blacktriangle}$} 	&	-290			&	-680			&	-290			&	520			\\  \cline{1-5}
			\multicolumn{1}{|c|}{$\zeta_{00\blacktriangle 0}$} 	&	-140			&	-50			&	160			&	130			\\ \cline{1-5}
			\multicolumn{1}{|c|}{$\zeta_{00\blacktriangle 1}$} 	&	-140			&	-50			&	100			&	110			\\  \cline{1-5}
			\multicolumn{1}{|c|}{$\zeta_{01\blacktriangle 0}$} 	&	-170			&	-50			&	160			&	100			\\  \cline{1-5}
			\multicolumn{1}{|c|}{$\zeta_{01\blacktriangle 1}$} 	&	-110			&	-50			&	160			&	70			\\ \cline{1-5}
			\multicolumn{1}{|c|}{$\zeta_{10\blacktriangle 0}$} 	&	-110			&	-50			&	160			&	40			\\  \cline{1-5}	
			\multicolumn{1}{|c|}{$\zeta_{10\blacktriangle 1}$} 	&	-50			&	-50			&	100			&	40			\\ \cline{1-5}
			\multicolumn{1}{|c|}{$\zeta_{11\blacktriangle 0}$} 	&	-170			&	-50			&	100			&	40			\\  \cline{1-5}
			\multicolumn{1}{|c|}{$\zeta_{11\blacktriangle 1}$} 	&	-140			&	-50			&	100			&	40			\\  \cline{1-5}
			\multicolumn{1}{|c|}{$\zeta_{0\blacktriangle 00}$} 	&	-485			&	-140			&	-50			&	-50			\\ \cline{1-5}
			\multicolumn{1}{|c|}{$\zeta_{0\blacktriangle 01}$} 	&	-470			&	-140			&	-50			&	-50			\\  \cline{1-5}
			\multicolumn{1}{|c|}{$\zeta_{0\blacktriangle 10}$} 	&	-485			&	-140			&	-50			&	-50			\\  \cline{1-5}
			\multicolumn{1}{|c|}{$\zeta_{0\blacktriangle 11}$} 	&	-470			&	-170			&	-50			&	-50			\\ \cline{1-5}
			\multicolumn{1}{|c|}{$\zeta_{1\blacktriangle 00}$} 	&	-215			&	-170			&	-50			&	-50			\\  \cline{1-5}	
			\multicolumn{1}{|c|}{$\zeta_{1\blacktriangle 01}$} 	&	-230			&	-140			&	-50			&	-50			\\ \cline{1-5}
			\multicolumn{1}{|c|}{$\zeta_{1\blacktriangle 10}$} 	&	-215			&	-110			&	-50			&	-50			\\  \cline{1-5}
			\multicolumn{1}{|c|}{$\zeta_{1\blacktriangle 11}$} 	&	-215			&	-110			&	-50			&	-50		\\  \cline{1-5}
		\end{tabular}
	\end{center}
	\caption{\label{table:STable1} \textbf{Frequencies obtained from Ramsey interferometry measurements} The $\blacktriangle$ indicates the qubit undergoing the Ramsey experiment in each row.}
\end{table}

\end{document}